# Using Correlation Adaptometry Method in Assessing Societal Stress: a Ukrainian Case

SVYATOSLAV R. RYBNIKOV,
Luhansk Institute of Interregional Academy of Personnel Management, Department of Social Sciences and Humanities
NATALIYA O. RYBNIKOVA,
Volodymyr Dahl East-Ukrainian National University, Department of Management and Economic Security

Societal stress may cause far reaching political, economic and even geological effects. Nevertheless, it is still scarcely investigated, contrary to social stress, which an individual faces in their interactions within a society. It is natural to suppose that in its adaptation to a stressor, society demonstrates the same objective laws that biological population does, since they are, in fact, the closest systems. In the survey, the hypothesis is tested that the collective stress effect holds true in society, which must appear – as it happens according to correlation adaptometry method in biological systems – in escalation of both correlations between societal characteristics and their dispersion. Both tends are observed in Ukrainian society during 2009-2012, as a result of political elections – the very priori stressor that affects societal anxiety.

## INTRODUCTION

Social stress, which an individual faces in their interactions within a society, is now a subject to a wide range of research, both physiological and psychological. To the contrary, societal stress, which a society feels as a whole, is still scarcely investigated, though it may have far ranging effects. It determines a country's foreign policy and military effort (Terrell, 1971); it provokes macroeconomic shifts – like it happened in Nigeria, where public anxiety concerning avian influenza stroked the rural economy (Fasina et al., 2010); it is considered to be even a geological force, able to trigger earthquakes (Molchanov, 2008).

Thus, it is crucial to assess societal stress, and to forecast its development in advance, which requires specific evaluation procedures. These procedures ought to be based on and complied with more general principles, e.g. ones of systemic methodology, and stress and coping theories. One successful example is the above mentioned research of Louis Terrell, where he confesses that he has borrowed key concepts from systems theory.

> "(...) to find the social and political conditions within a country which influence the levels of effort a nation is willing to expend on its military (...), it [this study] has borrowed notions from systems theory – stress and strain – which are represented empirically by indicators of societal frustration and social cleavage (stress) and of political instability (strain)" (Terrell, 1971: 344).

Nevertheless, in his paper Terrell uses rather physical than biological analogies – making an attempt to connect stress and strain quantitatively, like materials scientists commonly practice via their stress-strain curves. We find it more beneficial to borrow analogies from biological stress theory – in the form developed by Hans Selye and the followers. The reason is that biological and, especially, social systems are adaptive – contrary to physical systems that behave merely according to the laws of classical mechanics.

**AIM AND HYPOTHESIS**

In the theory of adaptation, the latter is considered as the process of a system's successful transition to a new functional state, as affected by factors that exert negative impact upon its normal functioning. In this context, "successful" means that in the functional state emerged the system renews its normal functioning, while negative factors endure. Alternatively, "unsuccessful" transition is one when the system looses its integrality and, therefore, is ruined in the process.

A system altering its functional state is considered to be a result of two groups of processes with contrary directions: adaptation and compensation, which tend to rebuild the system and to save its integrity while the rebuilding goes on. In this respect, a transition being successful or not, is determined by whether its compensatory mechanisms catch adaptation ones. These complex adaptation and compensation mechanisms carrying in a system while it is under rebuilding appear in a non-specific tension that Hans Selye would call "general adaptation syndrome", or later "stress" (Selye, 1936). The agents that do cause stress are naturally referred to as "stressors".

Assessing the level of a system's adaptation tension is of great practical importance. Data on its dynamics point out when a stressor occurred, and how the process develops. In 1987, Alexander Gorban et al. offered the correlation adaptometry method to assess the level of adaptation tension (Gorban, Manchuk, Petushkova, 1987). It is based on the collective stress effect that the researchers discovered before. The effect appears in escalation of: a) correlation between physiological parameters in population affected with a stressor; and b) their dispersion. In other words, stressors make different physiological parameters of a population's members change more concordantly, though within a wider range (in Figure 1, it is visualized in space of two physiological parameters).

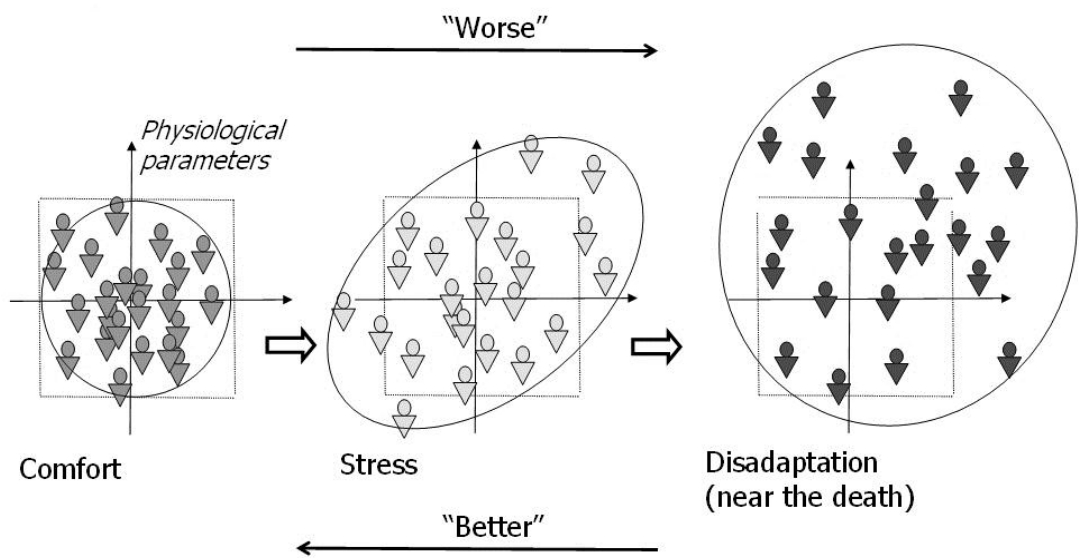

Figure 1. Collective stress effect (Gorban, Smirnova, Tyukina, 2010: 3194)

Discovered initially on the material of lipid metabolism parameters in the population of people working under the stress climatic conditions of Far North, the collective stress effect was later proved for other physiological parameters and even for other species' populations. Recently, the works appeared that expand the field of usage of the correlation adaptometry method to non-biological systems. Among the first researchers who, in fact, captured the collective stress effect in the field of finance were Françcois Longin and Bruno Solnik; in their case, the believed stressor was volatility – variation of market price:

> "We studied the correlation of monthly excess returns for seven major countries over the period 1960-90. (...) Tests of specific deviations lead to a rejection of the hypothesis of a constant conditional correlation. (...) We also find that the correlation rises in periods when the conditional volatility of markets is large. There is some preliminary evidence that economic variables such as the dividend yield and interest rates contain information about future volatility and correlation that is not contained in past returns alone" (Longin, Solnik, 1995: 19).

One of the recent works that intentionally try on the correlation adaptometry method to economic systems is (Gorban, Smirnova, Tyukina, 2010); there, financial crises are referred to as the adaptation tension in the economy.

It is worth to note that attempts to bring concepts from general and evolutionary ecology to economics are known for a long time. In the 1980s, they resulted in a pleiad of eminent works: Kenneth Boulding's *Evolutionary economics* (Boulding, 1981), Richard Nelson and Sidney Winter's *An evolutionary theory of economic change* (Nelson, Winter, 1982), and Michael Hannan and John Freeman's *The population ecology of organizations* (Hannan, Freeman, 1987), etc. They laid the foundations of a new area within economics, where economy is considered as a population of firms that emerge, develop and die like living things.

What is about society, it is, in fact, the nearest to biological population. Therefore, it is natural to suppose that the biological analogies are even more likely to work there. As Ralph Gerard, Clyde Kluckhohn, and Anatol Rapoport state,

> "Since Darwin there has been much discussion pro and con as to whether profitable analogies can be drawn between the evolution of species and (...) and of human societies. It is suggested that orientations and methods which have been employed to investigate biological evolution might also be used in the study of the evolution of society. Perhaps these will throw light on the theoretical problem of the similarities and differences in the two sorts of evolution" (Gerard, Kluckhohn, Rapoport, 1956: 6).

But it is also natural to check such assumptions, since "proof by analogy is not proof" – regardless of the fact, that "any similarity between two theories implies the existence of a more general theory of which the two are special cases" (Gerard, Kluckhohn, Rapoport, 1956: 7). According to Martin Sereno, an exhaustive validation must comprise of both theoretical and empirical procedures:

> "Complex generative and explanatory analogy is characterized by four distinct activities: 1) decomposition of the source and target systems, 2) establishment of a map between the two systems, 3) generation of predictions about the target, and 4) testing of the predictions" (Sereno, 1991: 468).

All the above concerns the collective stress effect, too. An appropriate theoretical validation of analogy between biological and social systems was given by Nicolas Rashevsky – a prominent Ukrainian-American scientist who founded both mathematical biology and mathematical sociology – in a series of works of 1960s on what he called "world set theory":

> "A society is a set of individuals plus the products of their activities, which result in their interactions. A multicellular organism is a set of cells plus the products of their activities, while a unicellular organism is a set of genes plus the products of their activities. (...) A physical system is a set of elementary particles plus the product of their activities, such as transitions from one energy level to another. Therefore physical, biological and sociological phenomena can be considered from a unified set-theoretical point of view" (Rashevsky, 1969: 159).

However, special experimental surveys, aimed to justify whether the collective stress effect holds true in society, are also required to complete the validation. Nevertheless, such research has not been conducted yet. In this respect, the article aims to answer whether using correlation adaptometry method is allowable in assessing societal stress. If so, it would help educe the tension early on and will be of considerable practical favor to social sciences.

**MATERIALS AND METHODS**

The input data were taken from the all-Ukrainian pools held by Rating Sociological Group in 2009-2012 (Rating, 2012). The question to the respondents was the following: "please, name three the most actual threats for Ukraine".

Their inquiries focused on fears of Ukrainians from 6 different regions, such as:
- West (Volyn, Zakarpattya, Ivano-Frankivsk, Rivne, Ternopil, Chernivtsi regions);
- Centre (Vinnytsya, Kyrovohrad, Poltava, Khmelnytsky, Cherkasy regions);
- North (Kyiv, Zhytomyr, Sumy, Chernihiv regions);
- South (Cremea, Odesa, Kherson, Mykolaiv regions);
- East (Dnipropetrovsk, Zaporizhzhya, Kharkiv regions);
- Donbas (Donetsk, Luhansk regions).

The inquiries' primary results are summarized in Appendix A.

Then, we suppose that the stressor that affects the society is elections and the changes in economic and social policy that follow them. Many social psychologists now agree that any election is a powerful stressor, due to "emotional pollution". Steven Sosny supposes that the latter is caused by two main factors: a) superabundance of false advertising during the campaign and b) politicians' being abnormally certain about enormously complex problems. Being a specialist in couples living in emotional abuse, he also names election one of the two conditions (together with economic crises) that greatly increase demand for services of family psychologists:

> "I specialize in couples living in resentment, anger, or emotional abuse. Two conditions greatly increase demand for my services: economic crises and national elections. When they occur together, it's like a perfect storm of family contention. I have written previously (...) about how so many people download and recycle the negativity in their environment and ultimately take it out on the closest people to them" (Stosny, 2008).

In their experimental study of 2011, Israeli researchers Israel Waismel-Manor, Gal Ifergane, and Hagit Cohen found that voters at the ballot box had higher positive and negative affect, as well exhibited cortisol levels that were significantly higher than their own normal levels obtained on a similar day. They think that election is a stressor due to the necessity of decision-making. Though in the experiment elevated cortisol levels calmed down just the next day, the researchers suppose it may have far-ranging social and societal outcomes, since such state is proved to affect memory consolidation, impair memory retrieval and lead to risk-seeking behavior (Waismel-Manor, Ifergan, Cohen, 2011).

Within the period analyzed, the presidential election was held in Ukraine (January-February 2010). We think that, in addition to common pre- and post-election psychological anxiety, an objective threat also occurred in this case. Namely, the out-party candidate became the winner. As a result, people had to adapt to a strong stressor – uncertainty concerning new economic and social policy.

Thus, if the collective stress effect does hold true in the society, and if the correlation adaptometry method does work, then increase in both correlation between fears and their dispersion must be fixed after the election. Later, as the society adapts to the stressor, these two characteristics are supposed to decrease again. And finally, we expect the second wave of societal stress, caused by upcoming parliamentary election, which took place in October 2012. The empirical material presented in Appendix A seems to be appropriate to test the hypothesis.

**Correlation Between Fears**

In (Gorban, Manchuk, and Petushkova, 1987), it is offered to depict the relations between the physiological parameters as a "correlation network" which vertices show the parameters considered, and edges connect pairs of the vertices with statistically strong interactions (edges closed to themselves, that indicate self-correlations, are not to be taken under consideration). Since each edge is attributed with the absolute value of a correlation coefficient, the network's total weight can be calculated, which represent the level of adaptation tension sought for.

According to this, to assess the adaptation tension of Ukrainian society in each of the periods analyzed, the correlation network's total weight was calculated, as the following:

$$w = \sum_{i=1}^{n-1} \sum_{j=i+1}^{n} |r_{ij}| \cdot z(r_{ij}), \qquad (1)$$

where $r_{ij}$ is Pearson's correlation coefficient between prevalence rates of fears $i$ and $j$.

The Pearson's correlation coefficient is defined as:

$$r_{ij} = \frac{\sum_{k=1}^{m} \left(x_{ik} - \overline{x_i^r}\right)\left(x_{jk} - \overline{x_j^r}\right)}{\sqrt{\sum_{k=1}^{m}\left(x_{ik} - \overline{x_i^r}\right)^2 \cdot \sum_{k=1}^{m}\left(x_{jk} - \overline{x_j^r}\right)^2}}, \qquad (2)$$

where $x_{ik}$, $x_{jk}$ are prevalence rates of fears $i$ and $j$, respectively, in region $k$; $\overline{x_i^r}$, $\overline{x_j^r}$ are average values of prevalence rates of fears $i$ and $j$, respectively, with respect to regions; $m$ is number of regions, and $z(r_{ij})$ is a correction factor, that excludes weak interactions out of the correlation network's total weight (only correlation coefficients which absolute values exceed a defined critical point are considered).

The correction factor is put conditionally, as:

$$z(r_{ij}) = \begin{cases} 1, & |r_{ij}| > r_0 \\ 0, & |r_{ij}| \le r_0 \end{cases}, \qquad (3)$$

where $r_0$ is the critical point (in this research $r_0$ equals 0.7).

**Dispersion of Fears**

Since there is no universally accepted measure for dispersion, we use two alternative estimates here in the study.

Formally, each region can be considered as a point in $n$-dimensional space, with the dimensions represented by the fears analyzed.

As an appropriate inferior estimate, we propose the diameter of $n$-dimensional ball with the volume equal to that of $n$-dimensional parallelepiped in which the points are submerged – since at a fixed volume ball is the solid with minimal linear size. The volume $V$ is the product of amplitudes:

$$V = \prod_{i=1}^{n} \left( \max_{k}\{x_{ik}\} - \min_{k}\{x_{ik}\} \right), \qquad (4)$$

where $x_{ik}$ is the prevalence rate of fear $i$ in region $k$; $n$ is number of fears.

In turn, the diameter $d_{min}$ sought can be obtained as:

$$d_{min} = \frac{2 \cdot \sqrt[n]{\Gamma\left(\frac{n}{2}+1\right)}}{\sqrt{\pi}} \cdot \sqrt[n]{V}, \qquad (5)$$

where $\Gamma(x)$ is Euler's gamma function.

As a superior estimate it is natural to use the maximal distance $d_{max}$ between the points:

$$d_{max} = \max_{k,l}\{d_{kl}\}. \qquad (6)$$

Here, the distance $d_{kl}$ between points $k$ and $l$ is defined as the Euclidean metric:

$$d_{kl} = \sqrt{\sum_{i=1}^{n}(x_{ik} - x_{il})^2}, \qquad (7)$$

## RESULTS AND DISCUSSION
### Correlation Between Fears

In Appendix B, the absolute values of the Pearson's correlation coefficients between prevalence rates of fears are shown (the cells for which $r_{ij} > 0.7$ are coloured light grey, and those for which $r_{ij} < -0.7$ are coloured dark grey).

In Figure 2, the dynamics of adaptation tension in Ukrainian society (calculated as the correlation network's total weight, or the sum of absolute values for colored cells from Appendix B) is shown.

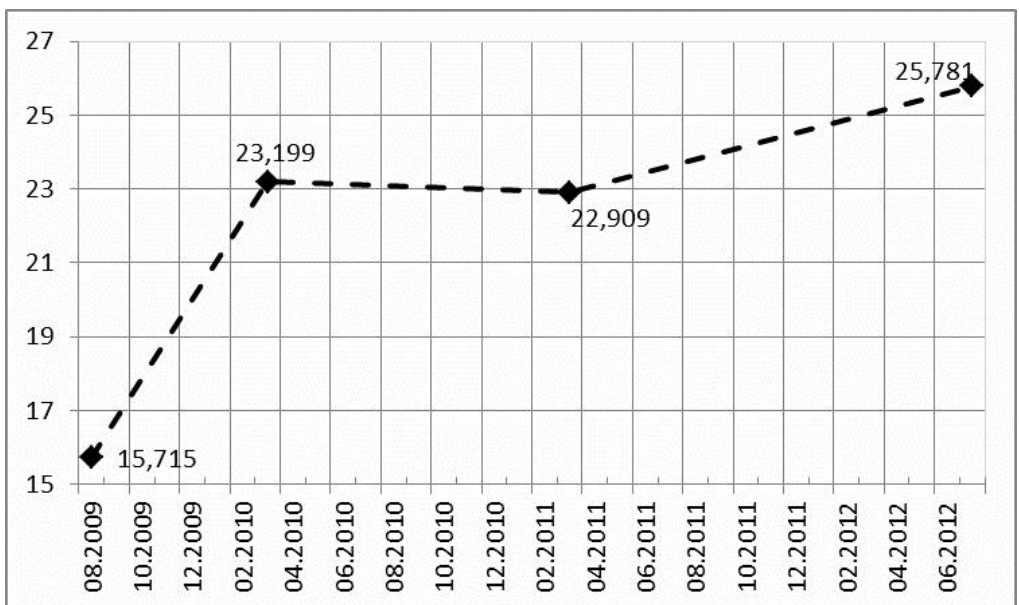

Figure 2. The dynamics of correlation between fears
(measured as the correlation network's total weight)

As one can see from the figure, in March 2010 the correlation network's total weight significantly rose comparing with August 2009 (23.20 as against 15.72 points). Then, in March 2011, it calmed down a bit (22.91 points), and in July 2012 demonstrated further growing (25.78 points). Thus, the assumptions above mentioned seem to have proved out in general.

The most questionable is the second interval, since the values of the correlation network's total weight in March 2011 and in March 2010 should hypothetically differ much more. We are inclined to seek the reason in the peculiarity of the initial data. Indeed, some fears are concerned to potential changes of political course after election, and they are naturally inherent to only a part of the society – namely, to respondents who approve of the alternative course. Thus, these fears can not characterize the adaptation tension of *the whole* society.

Speculatively, such fears are ones of military aggression from Russia (17) and military aggression from West (19). Alternatively, it is possible to analyze the initial data more formally, and to exclude fears with relatively high "political dispersion". In its 2012 survey, Rating Sociological Group examined not only geographical but also political distribution of fears (Rating, 2012). All the respondents were questioned which party they support. The answers were coded as following:
- A. "Support the Political Party "Yuliya Timoshenko Block"".
- B. "Support the Political Party "Svoboda"".
- C. "Support the Communist Party".
- D. "Support the Region Party".
- E. "Support the Political Party "Ukraina, vpered!"".
- F. "Support the Political Party "Udar"".
- G. "Do not support any political party".

For each fear, the coefficient of variation was calculated as a measure of dispersion, according to formula:

$$CV_i = \frac{\sqrt{\frac{\sum_{p=1}^{L}\left(x_{ip} - \overline{x_i^p}\right)^2}{L-1}}}{\overline{x_i^p}}, \tag{8}$$

where $x_{ip}$ is prevalence rate of *i* fear among the supporters of the political party $p$, and $\overline{x_i^p}$ is average value of prevalence rate of fear $i$ with respect to political party $p$.

In Table 1, the political distribution of fears is shown, according to (Rating, 2012), and the political dispersion of each fear calculated.

Table 1. Political distribution of fears in Ukrainian society, in 2012

| Code | Fear of | Prevalence rate of fear, %, by supporters of the political parties* | | | | | | | | CV |
|---|---|---|---|---|---|---|---|---|---|---|
| | | In general | A | B | C | D | E | F | G | |
| 1 | economic regress | 41 | 41 | 40 | 41 | 36 | 40 | 38 | 44 | 0.16 |
| 2 | rise in unemployment | 44 | 43 | 39 | 42 | 39 | 53 | 44 | 47 | 0.27 |
| 3 | depreciation of hryvnya | 15 | 19 | 12 | 12 | 12 | 17 | 24 | 13 | 0.77 |
| 4 | arbitrary rule | 25 | 32 | 30 | 23 | 10 | 25 | 29 | 26 | 0.72 |
| 5 | degeneracy of population | 20 | 18 | 30 | 29 | 21 | 24 | 18 | 17 | 0.73 |
| 6 | health services'worsening | 20 | 19 | 9 | 31 | 24 | 26 | 14 | 22 | 0.92 |
| 7 | environmental accidents | 17 | 12 | 15 | 14 | 21 | 13 | 15 | 18 | 0.51 |
| 8 | rise in crime | 16 | 17 | 7 | 16 | 17 | 15 | 13 | 17 | 0.61 |
| 9 | mass exodus | 10 | 10 | 7 | 9 | 10 | 13 | 15 | 9 | 0.67 |
| 10 | demographic crisis | 8 | 8 | 10 | 5 | 8 | 12 | 8 | 9 | 0.68 |

| Code | Fear of | Prevalence rate of fear, %, by supporters of the political parties* | | | | | | | | CV |
|---|---|---|---|---|---|---|---|---|---|---|
| | | In general | A | B | C | D | E | F | G | |
| 11 | schism of the state | 13 | 16 | 13 | 16 | 12 | 9 | 14 | 11 | 0.49 |
| 12 | loosing sovereignty | 11 | 17 | 33 | 5 | 7 | 4 | 11 | 11 | 2.27 |
| 13 | civil war | 5 | 5 | 12 | 6 | 6 | 6 | 5 | 5 | 1.44 |
| 14 | education services' worsening | 8 | 8 | 1 | 7 | 9 | 18 | 6 | 7 | 1.56 |
| 15 | loosing control over the gas-transport system | 5 | 6 | 3 | 1 | 5 | 6 | 6 | 5 | 0.96 |
| 16 | coup d'etat | 7 | 9 | 15 | 7 | 6 | 7 | 6 | 5 | 1.23 |
| 17 | military aggression from Russia | 3 | 5 | 9 | 0 | 1 | 1 | 4 | 2 | 2.56 |
| 18 | terrorism | 3 | 2 | 1 | 1 | 3 | 6 | 3 | 2 | 1.45 |
| 19 | military aggression from West | 1 | 1 | 3 | 2 | 3 | 4 | 1 | 1 | 4.24 |

* Letters refer to codes of the answers given above.

Thus, the most politically dispersed fears (in Table 3 they are colored grey) are the ones pointed out speculatively, i.e. fears of military aggression from Russia (17) and military aggression from West (19); these fears are believed to be the most appropriate candidates to exclude. In Figure 3, the dynamics of the correlation network's weight is shown built after excluding politically colored fears revealed. If we do exclude politically colored fears, the dynamics of the correlation network's weight does change, and responds to the hypothesis more accurately.

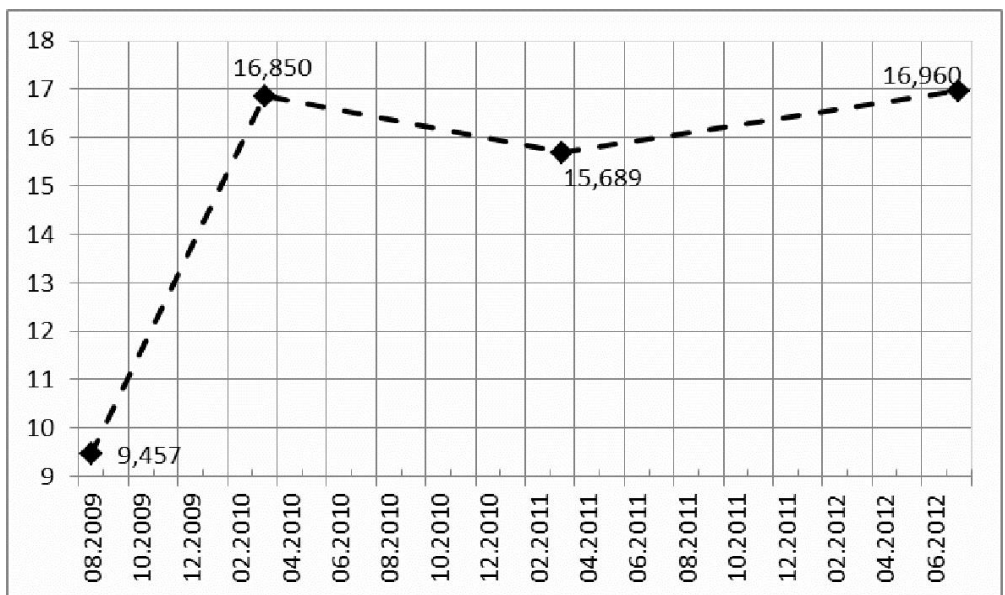

Figure 3. The dynamics of correlation between fears after excluding politically colored fears (measured as the correlation network's total weight)

**Dispersion of Fears**

In Table 2, the distances between regions in the "space" of fears analyzed are shown; they are calculated according to formula (7).
In Figure 4, the dynamics of two measures of dispersion are shown. It can be seen from the figure that it follows the dynamics of correlation between fears, in general.

Table 2. The "distances" between regions in the "space" of fears analyzed

| "Distance" between regions | West | Centre | North | East | South | Donbas |
|---|---|---|---|---|---|---|
| West: 2009 | 0 | 21.70 | 17.49 | 22.78 | 28.97 | 21.98 |
| 2010 | 0 | 26.63 | 34.80 | 46.16 | 33.03 | 39.69 |
| 2011 | 0 | 23.28 | 20.07 | 26.27 | 28.79 | 38.24 |
| 2012 | 0 | 38.52 | 25.71 | 30.50 | 40.72 | 40.27 |
| Centre: 2009 | | 0 | 26.63 | 27.75 | 31.59 | 23.24 |
| 2010 | | 0 | 20.30 | 28.39 | 15.75 | 25.88 |
| 2011 | | 0 | 21.19 | 24.58 | 19.72 | 24.98 |
| 2012 | | 0 | 26.48 | 27.24 | 30.79 | 19.54 |
| North: 2009 | | | 0 | 22.07 | 27.00 | 22.65 |
| 2010 | | | 0 | 32.03 | 24.58 | 25.50 |
| 2011 | | | 0 | 16.40 | 18.06 | 28.76 |
| 2012 | | | 0 | 13.96 | 33.39 | 24.06 |
| East: 2009 | | | | 0 | 14.90 | 10.49 |
| 2010 | | | | 0 | 19.65 | 35.97 |
| 2011 | | | | 0 | 11.87 | 24.66 |
| 2012 | | | | 0 | 34.55 | 22.09 |
| South: 2009 | | | | | 0 | 15.87 |
| 2010 | | | | | 0 | 21.68 |
| 2011 | | | | | 0 | 17.97 |
| 2012 | | | | | 0 | 28.18 |
| Donbas:2009 | | | | | | 0 |
| 2010 | | | | | | 0 |
| 2011 | | | | | | 0 |
| 2012 | | | | | | 0 |

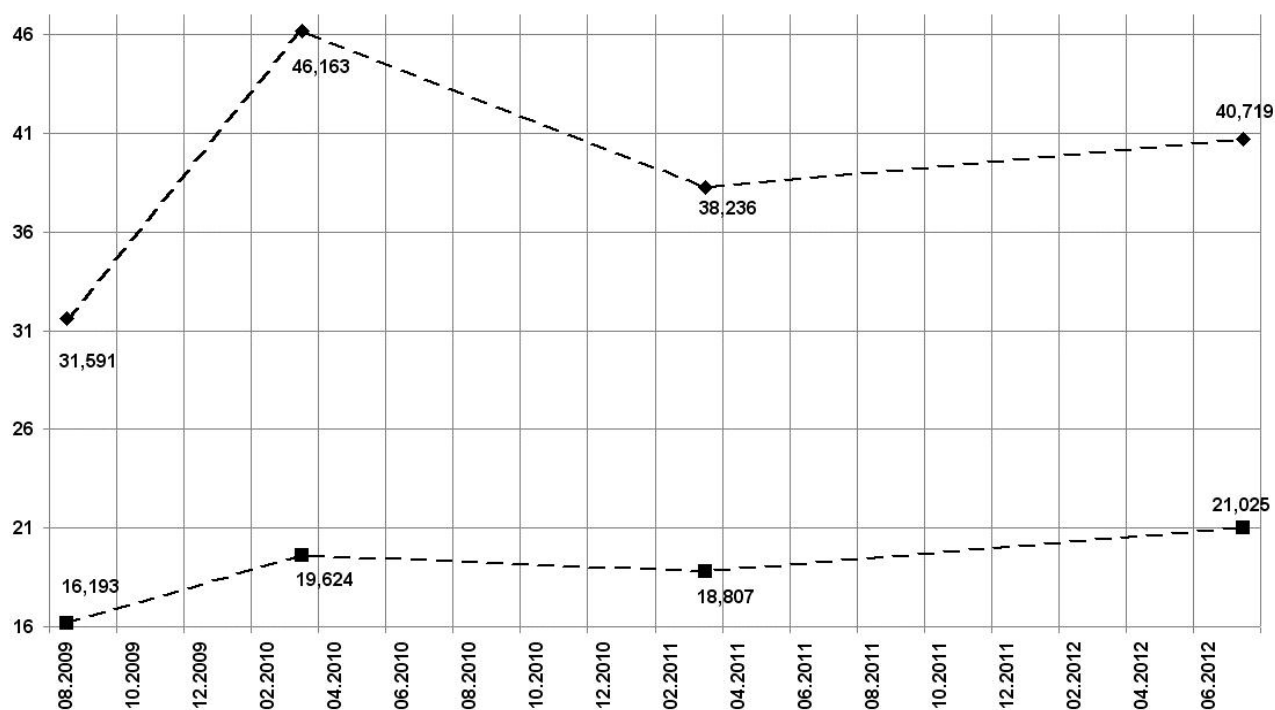

Figure 4. The dynamics of dispersion of fears

**Qualitative Analysis**

According to qualitative analysis undertaken, there are no pairs of fears which prevalence rates correlate statistically strongly during all the periods – neither positively nor negatively.

It can be also seen (Table 3) that, even though to exclude politically motivated fears (which are direct "markers" of the situation analyzed in Ukrainian society) there are some common, neutral fears which also signalize that the society is under stress. To such fears, we think it is naturally to subsume the fears of economic regress, degeneracy of population, and mass exodus. These fears' interactions with the others came to the fore, and the others began to correlate with these crucial ones.

It is also worth to note that such untypical interactions tend to attenuate during the next year (or two years). This could be interpreted as the renewal of normal functioning, which is the very same successful adaptation, considered at the beginning. But the fact of successful adaptation may be contravened by almost no decrease (in 2011 in comparison with 2010) and even increase (in 2012 in comparison with 2011) in general number of strong interactions between fears. We think that the increase in 2012 ought to be explained by the future stress which Ukrainian society has been looking for: at the end of October regular – parliamentary – elections will take place.

Table 3. Number of strong interactions of each fear with the others, by years

| Code | Fear of | 2009 | 2010 | 2011 | 2012 |
|---|---|---|---|---|---|
| 1 | economic regress | 3 | 7 | 7 | 2 |
| 2 | rise in unemployment | 1 | 2 | 1 | 2 |
| 3 | depreciation of hryvnya | 3 | 1 | 2 | 7 |
| 4 | arbitrary rule | 4 | 0 | 3 | 3 |
| 5 | degeneracy of population | 2 | 4 | 1 | 2 |
| 6 | health services' worsening | 1 | 1 | 3 | 4 |
| 7 | environmental accidents | 0 | 4 | 3 | 2 |
| 8 | rise in crime | 1 | 2 | 4 | 2 |
| 9 | mass exodus | 4 | 6 | 1 | 1 |
| 10 | demographic crisis | 2 | 2 | 1 | 2 |
| 11 | schism of the state | 2 | 5 | 3 | 6 |
| 12 | loosing sovereignty | 3 | 6 | 6 | 7 |
| 13 | civil war | 1 | 5 | 3 | 2 |
| 14 | education services' worsening | 1 | 0 | 2 | 5 |
| 15 | loosing control over the gas-transport system | 2 | 1 | 6 | 4 |
| 16 | coup d'etat | 2 | 4 | 4 | 5 |
| 18 | terrorism | 1 | 1 | 0 | 0 |
| Sum | | 33 | 51 | 50 | 56 |

The dynamics of both correlation between fears and their dispersion say that the collective stress effect holds true. Meanwhile, it is difficult to conclude whether the society has coped with the stress: in Figures 3 and 4, there are no clear-cut decreases after the election, in 2011, which would signal that the adaptation process has completed successfully. Instead, the re-increase observed in 2012, though slight, hints that the society is moving towards disadaptation state (in Figure 1, it is depicted right). Even if this scenario is realized, this should not be conceived apocalyptically – this will mean that fears are not be attributed to the regions in the way they do now.

**CONCLUSIONS**

Data on prevalence rates of fears reigning within the Ukrainian society push to the conclusion concerning the collective stress effect's holding true in social as well as in biological systems and, therefore, concerning the legacy of using correlation adaptometry method in assessing adaptation tension of societies. Testing the hypothesis was undertaken via "overlapping" the data on prevalence rates of fears and priory a strong stressor which is believed to be a political factor, e.g. switching governing teams – in January 2010 and in October 2012. And, really, proceeding from the calculations undertaken, the analyzed social system behaved as if it were a biological system reacting to the stress appeared: the dynamics of both correlations between the fears and dispersion of fears look like reflecting the system's behavior in the course of adaptation.

This pilot survey argue for the collective stress effect's holding true in social as well as in biological systems. Further research should imply involving vaster input data, describing the results of similar pools in Ukraine and other countries, both in past and in future. This is believed to result in revealing social systems which do or do not cope with the stress; analyzing structures of systems of these two classes would contribute to understanding, in terms of Martin Scheffer et al., "the architecture of fragility" (Scheffer et al., 2012). Involving vaster input data would also help select the most appropriate indicators to diagnose the situation more accurately, and, therefore, to choose the most valid signals for upcoming critical transitions.

## ACKNOWLEDGEMENT

We express deep appreciation to Prof. Alexander N. Gorban (Department of Mathematics, University of Leicester) for his valuable commentary in essence, and to Mrs. Yuliya Hassan for her kind style editing.

# APPENDICES

Appendix A. Geographical distribution of fears in Ukrainian society, by years, based on (Rating, 2012)

| Code | Fear of | Period | Prevalence rate of fear, %, by regions | | | | | |
|---|---|---|---|---|---|---|---|---|
| | | | West | Centre | North | East | South | Donbas |
| 1 | economic regress | 08.2009 | 54 | 65 | 51 | 62 | 60 | 64 |
| | | 03.2010 | 44 | 56 | 58 | 62 | 58 | 60 |
| | | 03.2011 | 43 | 56 | 49 | 50 | 57 | 67 |
| | | 07.2012 | 36 | 45 | 45 | 44 | 37 | 40 |
| 2 | rise in unemployment | 08.2009 | 44 | 43 | 35 | 46 | 46 | 44 |
| | | 03.2010 | 39 | 40 | 31 | 46 | 48 | 49 |
| | | 03.2011 | 46 | 47 | 44 | 52 | 50 | 44 |
| | | 07.2012 | 36 | 54 | 40 | 36 | 50 | 48 |
| 3 | depreciation of hryvnya | 08.2009 | 33 | 20 | 32 | 29 | 24 | 27 |
| | | 03.2010 | 29 | 20 | 19 | 29 | 25 | 20 |
| | | 03.2011 | 18 | 16 | 26 | 26 | 25 | 28 |
| | | 07.2012 | 21 | 15 | 19 | 13 | 12 | 10 |
| 4 | arbitrary rule | 08.2009 | 31 | 26 | 35 | 46 | 53 | 44 |
| | | 03.2010 | 20 | 26 | 26 | 46 | 31 | 15 |
| | | 03.2011 | 26 | 18 | 27 | 26 | 22 | 17 |
| | | 07.2012 | 28 | 20 | 20 | 19 | 42 | 20 |
| 5 | degeneracy of population | 08.2009 | 15 | 21 | 24 | 23 | 21 | 21 |
| | | 03.2010 | 10 | 16 | 23 | 23 | 14 | 19 |
| | | 03.2011 | 15 | 19 | 20 | 18 | 17 | 17 |
| | | 07.2012 | 18 | 18 | 22 | 20 | 18 | 28 |
| 6 | health services' worsening | 08.2009 | 12 | 11 | 13 | 10 | 16 | 11 |
| | | 03.2010 | 12 | 7 | 17 | 10 | 12 | 19 |
| | | 03.2011 | 9 | 15 | 17 | 17 | 19 | 17 |
| | | 07.2012 | 12 | 28 | 12 | 17 | 26 | 28 |
| 7 | environmental accidents | 08.2009 | 12 | 10 | 16 | 16 | 10 | 12 |
| | | 03.2010 | 7 | 11 | 22 | 16 | 15 | 24 |
| | | 03.2011 | 13 | 12 | 11 | 14 | 13 | 15 |
| | | 07.2012 | 15 | 11 | 18 | 22 | 16 | 21 |
| 8 | rise in crime | 08.2009 | 12 | 13 | 13 | 12 | 12 | 17 |
| | | 03.2010 | 14 | 15 | 15 | 12 | 13 | 14 |
| | | 03.2011 | 11 | 10 | 9 | 14 | 14 | 21 |
| | | 07.2012 | 10 | 13 | 17 | 12 | 28 | 18 |
| 9 | mass exodus | 08.2009 | 13 | 11 | 7 | 4 | 6 | 5 |
| | | 03.2010 | 12 | 7 | 6 | 4 | 5 | 6 |
| | | 03.2011 | 14 | 19 | 8 | 6 | 9 | 9 |
| | | 07.2012 | 13 | 16 | 8 | 6 | 7 | 11 |
| 10 | demographic crisis | 08.2009 | 6 | 5 | 11 | 4 | 6 | 6 |
| | | 03.2010 | 5 | 6 | 11 | 4 | 5 | 4 |
| | | 03.2011 | 9 | 13 | 13 | 6 | 6 | 9 |
| | | 07.2012 | 11 | 9 | 10 | 4 | 5 | 8 |
| 11 | schism of the state | 08.2009 | 6 | 14 | 9 | 8 | 14 | 9 |
| | | 03.2010 | 19 | 17 | 14 | 8 | 11 | 11 |
| | | 03.2011 | 8 | 14 | 7 | 5 | 8 | 14 |
| | | 07.2012 | 18 | 13 | 13 | 13 | 12 | 8 |
| 12 | loosing sovereignty | 08.2009 | 7 | 11 | 8 | 2 | 3 | 3 |
| | | 03.2010 | 26 | 9 | 10 | 2 | 5 | 4 |
| | | 03.2011 | 21 | 9 | 10 | 4 | 2 | 2 |
| | | 07.2012 | 28 | 5 | 11 | 9 | 6 | 4 |

| Code | Fear of | Period | Prevalence rate of fear, %, by regions | | | | | |
|---|---|---|---|---|---|---|---|---|
| | | | West | Centre | North | East | South | Donbas |
| 13 | civil war | 08.2009 | 3 | 4 | 4 | 3 | 5 | 7 |
| | | 03.2010 | 9 | 3 | 3 | 3 | 5 | 5 |
| | | 03.2011 | 9 | 8 | 9 | 6 | 10 | 4 |
| | | 07.2012 | 5 | 4 | 5 | 5 | 7 | 6 |
| 14 | education services' worsening | 08.2009 | 4 | 5 | 7 | 5 | 9 | 5 |
| | | 03.2010 | 5 | 2 | 5 | 5 | 5 | 3 |
| | | 03.2011 | 4 | 8 | 5 | 10 | 7 | 8 |
| | | 07.2012 | 4 | 8 | 8 | 11 | 8 | 9 |
| 15 | loosing control over the gas-transport system | 08.2009 | 9 | 5 | 12 | 10 | 11 | 6 |
| | | 03.2010 | 11 | 10 | 9 | 10 | 7 | 3 |
| | | 03.2011 | 9 | 7 | 10 | 4 | 5 | 2 |
| | | 07.2012 | 9 | 1 | 8 | 4 | 4 | 4 |
| 16 | coup d'etat | 08.2009 | 8 | 11 | 5 | 5 | 4 | 8 |
| | | 03.2010 | 10 | 8 | 11 | 5 | 7 | 5 |
| | | 03.2011 | 8 | 6 | 3 | 4 | 5 | 4 |
| | | 07.2012 | 13 | 6 | 5 | 4 | 8 | 4 |
| 17 | military aggression from Russia | 08.2009 | 5 | 5 | 5 | 2 | 1 | 0 |
| | | 03.2010 | 10 | 1 | 3 | 2 | 1 | 0 |
| | | 03.2011 | 6 | 2 | 3 | 3 | 0 | 1 |
| | | 07.2012 | 8 | 1 | 1 | 1 | 2 | 0 |
| 18 | terrorism | 08.2009 | 1 | 0 | 0 | 0 | 1 | 0 |
| | | 03.2010 | 0 | 0 | 1 | 0 | 1 | 0 |
| | | 03.2011 | 1 | 3 | 1 | 0 | 4 | 2 |
| | | 07.2012 | 2 | 3 | 3 | 2 | 3 | 2 |
| 19 | military aggression from West | 08.2009 | 1 | 3 | 1 | 1 | 2 | 1 |
| | | 03.2010 | 2 | 0 | 3 | 1 | 6 | 1 |
| | | 03.2011 | 1 | 2 | 1 | 1 | 1 | 2 |
| | | 07.2012 | 2 | 2 | 1 | 1 | 2 | 0 |

Population – people aged 18+.
Sample size – 2000.
Confidence level – 0,95.
Confidence intervals:
- 2,2% – for prevalence rates about 50%;
- 2,0% – for prevalence rates about 30%;
- 1,3% – for prevalence rates about 10%;
- 1,0% – for prevalence rates about 5%.

Appendix B1. Pearson's correlation coefficients between prevalence rates of fears (August 2009)

| Code of fear | 1 | 2 | 3 | 4 | 5 | 6 | 7 | 8 | 9 | 10 | 11 | 12 | 13 | 14 | 15 | 16 | 17 | 18 | 19 |
|---|---|---|---|---|---|---|---|---|---|---|---|---|---|---|---|---|---|---|---|
| 1 | 1.00 | | | | | | | | | | | | | | | | | | |
| 2 | 0.66 | 1.00 | | | | | | | | | | | | | | | | | |
| 3 | -0.79 | -0.37 | 1.00 | | | | | | | | | | | | | | | | |
| 4 | 0.19 | 0.44 | -0.02 | 1.00 | | | | | | | | | | | | | | | |
| 5 | 0.14 | -0.37 | -0.21 | 0.32 | 1.00 | | | | | | | | | | | | | | |
| 6 | -0.35 | -0.07 | -0.09 | 0.42 | -0.02 | 1.00 | | | | | | | | | | | | | |
| 7 | -0.46 | -0.46 | 0.68 | 0.07 | 0.48 | -0.37 | 1.00 | | | | | | | | | | | | |
| 8 | 0.38 | -0.08 | -0.11 | 0.07 | 0.14 | -0.30 | -0.10 | 1.00 | | | | | | | | | | | |
| 9 | -0.27 | -0.11 | 0.03 | -0.80 | -0.74 | -0.02 | -0.43 | -0.31 | 1.00 | | | | | | | | | | |
| 10 | -0.77 | -0.93 | 0.45 | -0.17 | 0.30 | 0.37 | 0.38 | 0.07 | 0.02 | 1.00 | | | | | | | | | |
| 11 | 0.49 | 0.13 | -0.90 | 0.13 | 0.33 | 0.46 | -0.62 | -0.06 | -0.07 | -0.13 | 1.00 | | | | | | | | |
| 12 | -0.20 | -0.53 | -0.19 | -0.91 | -0.15 | -0.10 | -0.22 | -0.16 | 0.76 | 0.32 | 0.22 | 1.00 | | | | | | | |
| 13 | 0.41 | 0.06 | -0.32 | 0.39 | 0.18 | 0.17 | -0.36 | 0.87 | -0.42 | 0.07 | 0.28 | -0.31 | 1.00 | | | | | | |
| 14 | -0.17 | -0.13 | -0.23 | 0.59 | 0.45 | 0.88 | -0.13 | -0.22 | -0.41 | 0.38 | 0.60 | -0.22 | 0.24 | 1.00 | | | | | |
| 15 | -0.73 | -0.32 | 0.58 | 0.42 | 0.25 | 0.54 | 0.54 | -0.55 | -0.29 | 0.51 | -0.22 | -0.31 | -0.37 | 0.58 | 1.00 | | | | |
| 16 | 0.42 | 0.06 | -0.41 | -0.76 | -0.39 | -0.53 | -0.48 | 0.32 | 0.63 | -0.30 | 0.12 | 0.65 | 0.07 | -0.67 | -0.90 | 1.00 | | | |
| 17 | -0.53 | -0.56 | 0.21 | -0.86 | -0.22 | -0.12 | 0.13 | -0.50 | 0.76 | 0.36 | -0.13 | 0.86 | -0.70 | -0.29 | 0.06 | 0.37 | 1.00 | | |
| 18 | -0.32 | 0.38 | 0.16 | 0.22 | -0.70 | 0.66 | -0.47 | -0.47 | 0.40 | -0.11 | 0.00 | -0.15 | -0.17 | 0.28 | 0.32 | -0.24 | 0.00 | 1.00 | |
| 19 | 0.51 | 0.17 | -0.90 | -0.29 | 0.04 | 0.17 | -0.70 | -0.18 | 0.34 | -0.30 | 0.87 | 0.54 | 0.00 | 0.20 | -0.47 | 0.50 | 0.21 | 0.00 | 1.00 |

Appendix B2. Pearson's correlation coefficients between prevalence rates of fears (March 2010)

| Code of fear | 1 | 2 | 3 | 4 | 5 | 6 | 7 | 8 | 9 | 10 | 11 | 12 | 13 | 14 | 15 | 16 | 17 | 18 | 19 |
|---|---|---|---|---|---|---|---|---|---|---|---|---|---|---|---|---|---|---|---|
| 1 | 1.00 | | | | | | | | | | | | | | | | | | |
| 2 | 0.36 | 1.00 | | | | | | | | | | | | | | | | | |
| 3 | -0.36 | 0.29 | 1.00 | | | | | | | | | | | | | | | | |
| 4 | 0.46 | 0.14 | 0.49 | 1.00 | | | | | | | | | | | | | | | |
| 5 | 0.81 | -0.13 | -0.34 | 0.44 | 1.00 | | | | | | | | | | | | | | |
| 6 | 0.17 | -0.03 | -0.40 | -0.52 | 0.30 | 1.00 | | | | | | | | | | | | | |
| 7 | 0.73 | 0.10 | -0.59 | -0.11 | 0.74 | 0.78 | 1.00 | | | | | | | | | | | | |
| 8 | -0.31 | -0.67 | -0.75 | -0.70 | -0.12 | 0.19 | 0.05 | 1.00 | | | | | | | | | | | |
| 9 | -0.98 | -0.36 | 0.25 | -0.58 | -0.75 | -0.07 | -0.64 | 0.41 | 1.00 | | | | | | | | | | |
| 10 | 0.00 | -0.88 | -0.55 | -0.11 | 0.36 | 0.27 | 0.28 | 0.64 | -0.01 | 1.00 | | | | | | | | | |
| 11 | -0.86 | -0.59 | -0.10 | -0.57 | -0.65 | -0.23 | -0.61 | 0.72 | 0.87 | 0.28 | 1.00 | | | | | | | | |
| 12 | -0.98 | -0.50 | 0.31 | -0.45 | -0.68 | -0.09 | -0.65 | 0.38 | 0.97 | 0.14 | 0.86 | 1.00 | | | | | | | |
| 13 | -0.85 | 0.08 | 0.50 | -0.49 | -0.81 | 0.11 | -0.51 | -0.02 | 0.83 | -0.33 | 0.51 | 0.79 | 1.00 | | | | | | |
| 14 | -0.15 | -0.16 | 0.60 | 0.40 | 0.04 | 0.17 | -0.04 | -0.49 | 0.02 | 0.18 | -0.23 | 0.20 | 0.28 | 1.00 | | | | | |
| 15 | -0.48 | -0.57 | 0.46 | 0.44 | -0.17 | -0.70 | -0.74 | 0.02 | 0.38 | 0.24 | 0.47 | 0.52 | 0.08 | 0.29 | 1.00 | | | | |
| 16 | -0.62 | -0.91 | -0.17 | -0.31 | -0.25 | 0.03 | -0.29 | 0.66 | 0.58 | 0.78 | 0.75 | 0.71 | 0.25 | 0.26 | 0.53 | 1.00 | | | |
| 17 | -0.91 | -0.43 | 0.56 | -0.19 | -0.56 | -0.10 | -0.64 | 0.09 | 0.87 | 0.06 | 0.65 | 0.94 | 0.79 | 0.46 | 0.60 | 0.60 | 1.00 | | |
| 18 | 0.20 | -0.30 | -0.28 | 0.08 | 0.15 | 0.29 | 0.32 | 0.11 | -0.32 | 0.64 | -0.16 | -0.16 | -0.22 | 0.49 | -0.09 | 0.41 | -0.18 | 1.00 | |
| 19 | 0.00 | 0.08 | 0.13 | 0.08 | -0.23 | 0.19 | 0.06 | -0.23 | -0.16 | 0.18 | -0.19 | -0.04 | 0.17 | 0.62 | -0.14 | 0.20 | 0.00 | 0.85 | 1.00 |

Appendix B3. Pearson's correlation coefficients between prevalence rates of fears (March 2011)

| Code of fear | 1 | 2 | 3 | 4 | 5 | 6 | 7 | 8 | 9 | 10 | 11 | 12 | 13 | 14 | 15 | 16 | 17 | 18 | 19 |
|---|---|---|---|---|---|---|---|---|---|---|---|---|---|---|---|---|---|---|---|
| 1 | 1.00 | | | | | | | | | | | | | | | | | | |
| 2 | -0.18 | 1.00 | | | | | | | | | | | | | | | | | |
| 3 | 0.42 | 0.02 | 1.00 | | | | | | | | | | | | | | | | |
| 4 | -0.87 | 0.22 | 0.09 | 1.00 | | | | | | | | | | | | | | | |
| 5 | 0.11 | -0.09 | 0.17 | 0.01 | 1.00 | | | | | | | | | | | | | | |
| 6 | 0.60 | 0.27 | 0.69 | -0.24 | 0.57 | 1.00 | | | | | | | | | | | | | |
| 7 | 0.51 | 0.22 | 0.40 | -0.39 | -0.57 | 0.08 | 1.00 | | | | | | | | | | | | |
| 8 | 0.77 | -0.03 | 0.60 | -0.55 | -0.36 | 0.33 | 0.91 | 1.00 | | | | | | | | | | | |
| 9 | -0.07 | -0.24 | -0.92 | -0.42 | -0.08 | -0.54 | -0.32 | -0.39 | 1.00 | | | | | | | | | | |
| 10 | -0.06 | -0.71 | -0.41 | -0.14 | 0.57 | -0.21 | -0.68 | -0.50 | 0.55 | 1.00 | | | | | | | | | |
| 11 | 0.70 | -0.51 | -0.27 | -0.93 | 0.02 | -0.02 | 0.19 | 0.36 | 0.61 | 0.44 | 1.00 | | | | | | | | |
| 12 | -0.77 | -0.33 | -0.67 | 0.46 | -0.35 | -0.93 | -0.41 | -0.60 | 0.49 | 0.37 | -0.15 | 1.00 | | | | | | | |
| 13 | -0.59 | 0.12 | -0.43 | 0.45 | 0.02 | -0.17 | -0.75 | -0.77 | 0.25 | 0.13 | -0.38 | 0.43 | 1.00 | | | | | | |
| 14 | 0.51 | 0.59 | 0.30 | -0.40 | 0.26 | 0.57 | 0.52 | 0.46 | -0.23 | -0.38 | 0.12 | -0.76 | -0.61 | 1.00 | | | | | |
| 15 | -0.77 | -0.33 | -0.51 | 0.59 | 0.20 | -0.50 | -0.88 | -0.90 | 0.34 | 0.60 | -0.30 | 0.78 | 0.74 | -0.78 | 1.00 | | | | |
| 16 | -0.40 | 0.03 | -0.82 | -0.03 | -0.70 | -0.83 | 0.00 | -0.26 | 0.70 | -0.04 | 0.15 | 0.71 | 0.35 | -0.41 | 0.29 | 1.00 | | | |
| 17 | -0.83 | -0.16 | -0.46 | 0.62 | -0.33 | -0.88 | -0.20 | -0.48 | 0.21 | 0.18 | -0.36 | 0.91 | 0.17 | -0.53 | 0.61 | 0.54 | 1.00 | | |
| 18 | 0.52 | 0.01 | -0.19 | -0.63 | -0.03 | 0.34 | -0.10 | 0.10 | 0.39 | 0.01 | 0.52 | -0.34 | 0.34 | 0.00 | -0.17 | 0.15 | -0.69 | 1.00 | |
| 19 | 0.73 | -0.40 | -0.18 | -0.92 | 0.15 | 0.07 | 0.27 | 0.42 | 0.51 | 0.41 | 0.96 | -0.27 | -0.57 | 0.35 | -0.42 | 0.00 | -0.37 | 0.35 | 1.00 |

Appendix B4. Pearson's correlation coefficients between prevalence rates of fears (July 2012)

| Code of fear | 1 | 2 | 3 | 4 | 5 | 6 | 7 | 8 | 9 | 10 | 11 | 12 | 13 | 14 | 15 | 16 | 17 | 18 | 19 |
|---|---|---|---|---|---|---|---|---|---|---|---|---|---|---|---|---|---|---|---|
| 1 | 1.00 | | | | | | | | | | | | | | | | | | |
| 2 | 0.08 | 1.00 | | | | | | | | | | | | | | | | | |
| 3 | -0.03 | -0.53 | 1.00 | | | | | | | | | | | | | | | | |
| 4 | -0.74 | 0.22 | -0.08 | 1.00 | | | | | | | | | | | | | | | |
| 5 | 0.12 | 0.05 | -0.46 | -0.45 | 1.00 | | | | | | | | | | | | | | |
| 6 | -0.01 | 0.90 | -0.80 | 0.16 | 0.22 | 1.00 | | | | | | | | | | | | | |
| 7 | 0.06 | -0.50 | -0.41 | -0.26 | 0.64 | -0.19 | 1.00 | | | | | | | | | | | | |
| 8 | -0.31 | 0.51 | -0.51 | 0.72 | 0.08 | 0.46 | 0.05 | 1.00 | | | | | | | | | | | |
| 9 | 0.00 | 0.42 | 0.27 | -0.24 | -0.11 | 0.28 | -0.73 | -0.45 | 1.00 | | | | | | | | | | |
| 10 | -0.05 | -0.07 | 0.73 | -0.26 | 0.05 | -0.35 | -0.47 | -0.43 | 0.67 | 1.00 | | | | | | | | | |
| 11 | -0.24 | -0.54 | 0.86 | 0.16 | -0.75 | -0.69 | -0.44 | -0.50 | 0.21 | 0.40 | 1.00 | | | | | | | | |
| 12 | -0.48 | -0.69 | 0.82 | 0.10 | -0.37 | -0.76 | -0.17 | -0.51 | 0.18 | 0.54 | 0.87 | 1.00 | | | | | | | |
| 13 | -0.63 | 0.15 | -0.50 | 0.74 | 0.23 | 0.28 | 0.36 | 0.85 | -0.57 | -0.46 | -0.40 | -0.22 | 1.00 | | | | | | |
| 14 | 0.60 | 0.14 | -0.72 | -0.34 | 0.36 | 0.35 | 0.58 | 0.19 | -0.52 | -0.76 | -0.69 | -0.79 | 0.08 | 1.00 | | | | | |
| 15 | -0.36 | -0.75 | 0.73 | 0.08 | 0.00 | -0.86 | 0.18 | -0.20 | -0.21 | 0.51 | 0.55 | 0.80 | 0.07 | -0.59 | 1.00 | | | | |
| 16 | -0.74 | -0.24 | 0.63 | 0.53 | -0.57 | -0.36 | -0.51 | -0.15 | 0.34 | 0.45 | 0.80 | 0.84 | 0.04 | -0.92 | 0.53 | 1.00 | | | |
| 17 | -0.66 | -0.48 | 0.71 | 0.33 | -0.50 | -0.54 | -0.34 | -0.36 | 0.28 | 0.47 | 0.86 | 0.95 | -0.09 | -0.87 | 0.64 | 0.96 | 1.00 | | |
| 18 | 0.31 | 0.57 | 0.09 | 0.30 | -0.37 | 0.21 | -0.58 | 0.51 | 0.05 | 0.07 | -0.06 | -0.39 | 0.00 | 0.00 | -0.25 | -0.11 | -0.31 | 1.00 | |
| 19 | -0.26 | 0.13 | 0.46 | 0.55 | -0.96 | -0.09 | -0.80 | 0.01 | 0.30 | 0.12 | 0.72 | 0.38 | -0.16 | -0.54 | 0.00 | 0.69 | 0.56 | 0.45 | 1.00 |

# AUTHOR BIOGRAPHIES

SVYATOSLAV R. RYBNIKOV

MSc in Ecology and Environmental Protection, gained at National University of "Kyiv-Mohyla Academy" (Kyiv, Ukraine) in 2004.

PhD in Theory and Methods of Professional Education (teaching environmental management), gained at Taras Shevchenko Luhansk National University (Luhansk, Ukraine) in 2011.

Present affiliation: Luhansk Institute of Interregional Academy of Personnel Management, Department of Social Sciences and Humanities.

Research interests: adapting concepts from natural sciences, especially ecological and evolutionary ones, to economic and social systems.

E-mail: rybnikov@ukr.net

NATALIYA O. RYBNIKOVA

MSc in Management, gained at Volodymyr Dahl East-Ukrainian National University (Luhansk, Ukraine) in 2003.

PhD in Enterprise Economic Security, gained at Volodymyr Dahl East-Ukrainian National University (Luhansk, Ukraine) in 2011.

Present affiliation: Volodymyr Dahl East-Ukrainian National University, Department of Management and Economic Security.

Research interests: adapting concepts from natural sciences, especially ecological and evolutionary ones, to economic and social systems.

E-mail: rybnikova@ukr.net